\documentstyle[12pt,aaspp]{article}
\input psfig.sty

\begin{document}
\def\M        {{$-$}}
\def\VT       {{$V_3$}}
\def\hh       {{$^h$}}
\def\mm       {{$^m$}}
\def\ss       {{$^s$}}                  
\def\deg      {{\ifmmode^\circ\else$^\circ$\fi}}  
\def\degree   {{\ifmmode^\circ\else$^\circ$\fi} } 
\def\arcm     {{\ifmmode {'  }\else$'     $\fi}}  
\def\arcmin   {{\ifmmode {'  }\else$'     $\fi} } 
\def\arcs     {{\ifmmode {'' }\else$''    $\fi}}  
\def\arcsec   {{\ifmmode {'' }\else$''    $\fi} } 
\def\secs     {{$^{\prime\prime}$}}               
\def\sspt     {{$\buildrel{s}           \over .$}}
\def\degpt    {{$\buildrel{\circ}       \over .$}}
\def\adegpt   {{$\buildrel{\circ}       \over .$}}
\def\arcmpt   {{$\buildrel{\prime}      \over .$}}
\def\arcspt   {{$\buildrel{\prime\prime}\over .$}}
\def\adeg     {{\hskip .10em {}^{\circ}\hskip -.40em . \hskip.25em}}
\def\amin     {{\hskip .10em {}'       \hskip -.30em . \hskip.20em}}
\def\asec     {{\hskip .10em {}''      \hskip -.49em . \hskip.29em}}

\def\magarc   {{mag\ arcsec$^{-2}$}}
\def\et       {{et\thinspace al.} }	
\def\etal     {{et\thinspace al.} }	
\def\eg       {{\it e.g.}, }
\def\ie       {{\it i.e.}, }
\def\chisq    {{$\chi^{2}$}}

\def\apgt{\ {\raise-.5ex\hbox{$\buildrel>\over\sim$}}\ }
\def\aplt{\ {\raise-.5ex\hbox{$\buildrel<\over\sim$}}\ }

\def\Iband {\hbox{$I_{814}$}}
\def\Vband {\hbox{$V_{606}$}}
\def\magsq {\hbox{$\,\rm mag~arcsec^{-2}$}}
\def\kms   {\hbox{$\,\rm km~s^{-1}$}}


\title{NEW ``EINSTEIN CROSS'' GRAVITATIONAL LENS \\
  CANDIDATES IN HST WFPC2 SURVEY IMAGES}
\author{Kavan U. Ratnatunga, Eric J. Ostrander, Richard E. Griffiths, \& Myungshin Im}
\affil{Bloomberg Center for Physics and Astronomy, \\
Johns Hopkins University, Baltimore, MD 21218 \\
 kavan, ejo, myung, \& griffith@mds.pha.jhu.edu}

{\it To be published Astrophysical Journal Letters of November 1 1995 }

{\it Received June 26 1995 \hskip 0.5in Accepted Aug 24 1995} 

\begin{abstract}

We report the serendipitous discovery of ``Einstein cross''
gravitational lens candidates using the Hubble Space Telescope.  We have
so far discovered two good examples of such lenses, each in the form of
four faint blue images located in a symmetric configuration around a red
elliptical galaxy. The high resolution of HST has facilitated the
discovery of this optically selected sample of faint lenses with small
($\sim$1\arcsec) separations between the (I$\sim25-27$) lensed
components and the much brighter (I$\sim19-22$) lensing galaxies. The
sample has been discovered in the routine processing of HST fields
through the Medium Deep Survey pipeline, which fits simple galaxy models
to broad band filter images of all objects detected in random survey
fields using WFPC2.

We show that the lens configuration can be modeled using the
gravitational field potential of a singular isothermal ellipsoidal mass
distribution.  With this model the lensing potential is very similar,
both in ellipticity and orientation, to the observed light distribution
of the elliptical galaxy, as would occur when stars are a tracer
population.  The model parameters and associated errors have been
derived by 2-dimensional analysis of the observed images. The maximum
likelihood procedure iteratively converges simultaneously on the model
for the lensing elliptical galaxy and the source of the lensed
components.  A systematic search is in progress for other gravitational
lens candidates in the HST Medium Deep Survey.  This should eventually
lead to a good statistical estimate for lensing probabilities, and
enable us to probe the cosmological component of the observed faint
blue galaxy population.
 
\end{abstract}

\clearpage
\section {INTRODUCTION}

Einstein (1936) computed the gravitational deflection of light by
massive objects and showed that an image can be highly magnified if the
observer, source and the deflector are sufficiently well
aligned. However, the angular resolution available then to ground based
optical telescopes made him remark that ``there is no great chance of
observing this phenomenon''.  Zwicky (1937) showed that ``extragalactic
{\it nebulae} offer a much better chance than {\it stars} for the
observation of gravitational lens effects''.

Over the last decade a number of lensed QSO candidates were located in
radio surveys and subsequently the associated lensing galaxies were
optically identified (See Schneider, Ehlers and Falco 1992 for
review). Huchra \etal (1985) discovered the ``Einstein cross'' at the
center of the bright (V=14.6) galaxy 2237+0305, an object in the Center
for Astrophysics redshift survey: this lens is considered unique because
of the very low probability of alignment of a QSO within 0\farcs3 of the
center of a nearby (z=0.04) galaxy.  With the refurbished WFPC2 optics
on the Hubble Space Telescope (HST), we have now been able to start an
optical survey for gravitational lenses centered on field galaxies in
the magnitude range I$ = 19 - 23$, searching for background field
galaxies which are lensed into components with magnitudes I$ = 23 - 27$.

The Medium Deep Survey (MDS) is an HST key project which relies
exclusively on the efficient use of parallel observing time to take
WFPC2 images of random fields which are several arcminutes away from the
primary targets of other HST instruments (Griffiths \etal 1994).
Similar observations have been made by the Guaranteed Time Observers
(GTO) using WFPC2 in parallel mode, in conjunction with primary GTO
exposures. The GTO parallel observations have been made available to all
HST GTO teams. In addition to these two parallel surveys, a major
`strip' survey was performed (Groth \etal 1995) using the WFPC2 in
primary mode.

For the study of galaxies in the parallel surveys (MDS and GTO), we
exclude fields at low galactic latitude and those inside Galactic
globular clusters or local group galaxies, where the data are dominated
by stellar images. The Groth-Westphal `strip' and other fields with both
I and V observations are processed through the MDS pipeline as they
become available via the HST archive after the one-year proprietary
period.  We fit simple galaxy models to images of all extended objects
detected, using the maximum likelihood method for determination of the
best-fit galaxy parameters (Ratnatunga \etal 1995).  In view of the
large number of fields, most of the processing is automated except for a
manual stage to correct for errors in the detection and resolution of
objects, and in the final inspection of the fitted models and residual
images.

Serendipitous discoveries have always been a major potential of the
Medium Deep Survey. Previous results from the MDS have focussed on the
systematic measurements of stars and galaxies (see Griffiths et al 1995
for a summary and a bibliography for the MDS).  Although an unusual
object was discovered in the very first MDS image from WFPC2 (Glazebrook
et al. 1994), this was probably a proto-galaxy in formation at z = 0.7
with superluminous starburst knots, rather than a gravitational lens as
originally suspected.  The observations presented in this letter
represent the first discovery of a relatively new class of objects.

\section {OBSERVATIONS}

Prior to December 1993, the aberrated data from HST and WF/PC could not
be used for the purposes of the present study.  In cycle 4 of HST
observations, from January 1994 to June 1995, the MDS and GTO parallel
datasets have comprised about 35 and 15 independent high galactic
latitude fields, respectively, with at least two WFPC2 exposures in each
of the F606W(V) and F814W(I) filters for each field.  These fields have
all been processed through the MDS pipeline. In addition, there are
about 30 and 70 fields in these surveys, respectively, for which one or
both of the filters has only one exposure; in these latter fields,
unbiased cosmic ray cleaning is a more difficult task, and processing of
these fields is not yet completed.

The 42 arc min long Groth-Westphal strip (Groth \etal 1995) consists of
28 contiguous WFPC2 fields centered at $b=$+60\degpt25 and
$l=$96\degpt35. The observations, covering a total area about 120 square
arc minutes, were taken between 7 March and 9 April 1994. They were
obtained from the HST archive and calibrated, stacked and processed in
exactly the same way as fields obtained for the HST Medium Deep
Survey. A root-mean-square(rms) error image reflecting both the excluded
cosmic rays and the flat field was created and used in the object
detection and subsequent image analysis algorithms.  Full details of the
catalog and statistical analysis of the distributions of magnitude,
color, half-light radius, axis ratio, and position angle and location
will be described in other publications.

In order to search for serendipitous objects and to correct any errors
in the automated object detection process caused by confusion of
overlapping or very bright images, the fields have been examined by one
of us (EJO) by eye. During this process, it was noticed that the I=19.7
elliptical galaxy (HST14176+5226) was flanked by four fainter images
which were all at about the same magnitude and color and much bluer than
the central elliptical which had a half light radius of 1\farcs2 (see
Figure 1).  The companion objects (V$\sim26$) were about 1\farcs2 and
1\farcs6 distant from the center of the elliptical galaxy along the
major and minor axes respectively.  Furthermore, the objects on the
cross appeared to be arcs. This galaxy was thus clearly identified as 
a very strong gravitational lens candidate.

\begin{figure}[tb!]
\psfig{file=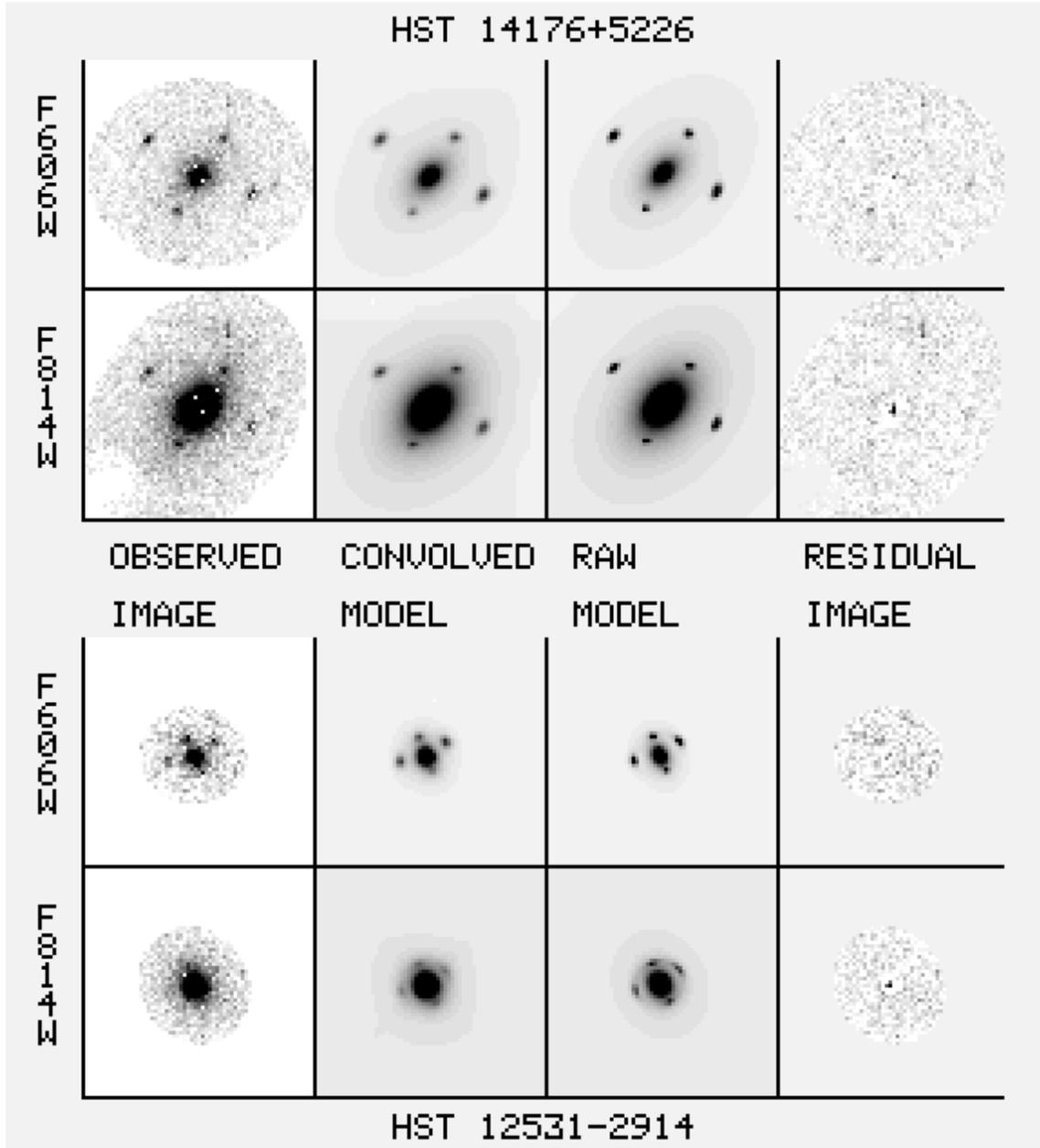,width=6.0in}
\caption{The maximum likelihood fits to the two gravitationally lensed
images.  The images are displayed as analysed, without any interpolation
over bad pixels. Each box is 6\farcs4 square. The residuals show only a
very faint trace of the subtracted images. Note that the convolution
with the WFPC2 PSF does influence the appearance of the lensed
images.}
\end{figure} 

A second fainter (I=21.8) and smaller (half light radius 0\farcs2)
elliptical galaxy (HST12531\M2914) was discovered by one of us (MI) when
inspecting the residuals of the maximum likelihood model fits to
galaxies in a deep MDS field urz00.  Since the faint companion objects
(V$\sim27$) are only about 0\farcs5 from the central elliptical, they
had not been resolved as separate objects by the automated object
detection algorithm.

The observed image configuration, magnitudes and colors are given at the
top of table~1. The magnitudes of the components were computed using a
0\farcs3 square aperture, and corrected to total magnitudes assuming a
point source. The magnitude of the elliptical is from the model fit. 
The offsets (X,Y) are in WFC 0\farcs1 pixels from the
centroid of the respective elliptical galaxy.

\begin{table*}[tb!]
\begin{center}
\caption{HST `Einstein Cross' Gravitational Lens Candidates}
\begin{tabular}{l | r r c c c c | r r c c c c }
\tableline
 Name                & \multicolumn{6}{c|}{HST14176+5226}                                 & \multicolumn{6}{c}{HST12531\M 2914} \\
 Equ(J2000)  & \multicolumn{2}{c}{14:17:36.3} & \multicolumn{2}{c}{+52:26:44} & \multicolumn{2}{c|}{\VT=32\degpt93} & \multicolumn{2}{c}{12:53:06.7} & \multicolumn{2}{c}{\M 29:14:30} & \multicolumn{2}{c}{\VT=127\degpt77}\\
 HST WFPC2  & \multicolumn{6}{c|}{Groth GTO:5090 11-Mar-1994}                        & \multicolumn{6}{c}{Griffiths GO:5369 15-Feb-1995} \\
 Dataset[g][x,y] & \multicolumn{6}{c|}{U26X0801T[3][242,700]}                         & \multicolumn{6}{c}{U26K7G04T[3][755,326]} \\ \hline
 Configuration &  X  & Y  &   V   & $\pm$ & V\M I  & $\pm$ & X  & Y  &   V   & $\pm$ & V\M I  & $\pm$  \\
 Elliptical &   0 &  0 & 21.68 & 0.04  & 1.97 & 0.04  &  0 &  0 & 23.77 & 0.06  & 1.95 & 0.07   \\   
    A       & \M 11 & 11 & 25.63 & 0.06  & 0.51 & 0.10  & \M 6 & \M 2 & 27.02 & 0.15  & 0.25 & 0.28   \\   
    B       &  18 & \M 4 & 25.77 & 0.07  & 0.39 & 0.11  &  6 &  3 & 26.89 & 0.15  & 0.43 & 0.23   \\   
    C       &  10 & 12 & 25.99 & 0.08  & 0.52 & 0.13  & \M 2 &  4 & 26.72 & 0.11  & 0.34 & 0.21   \\   
    D       &  \M 3 & \M 9 & 25.97 & 0.08  & 0.42 & 0.14  &  3 & \M 4 & 27.51 & 0.24  & 0.82 & 0.34   \\   
\end{tabular}
\begin{tabular}{l | c l c l | c l c l }
\tableline
 HST WFPC2 Filter    & \multicolumn{2}{c}{F814W}    & \multicolumn{2}{c|}{F606W}          & \multicolumn{2}{c}{F814W}    & \multicolumn{2}{c}{F606W} \\
 Exposure seconds    & \multicolumn{2}{c}{4 x 1100} & \multicolumn{2}{c|}{4 x 700}        & \multicolumn{2}{c}{4 x 2100} & \multicolumn{2}{c}{3 x 1800} \\
 Fitted Parameter             &  MLE & $\pm$ rms &  MLE & $\pm$ rms &  MLE & $\pm$ rms &  MLE & $\pm$ rms \\
 Sky Mag. ($arcsec^{2}$)      &  22.281 & 0.003 &  22.807 & 0.003 &  21.950 & 0.004 &  22.517 & 0.004 \\
 Total Mag of elliptical      &  19.71  & 0.01  &  21.68  & 0.04  &  21.82  & 0.03  &  23.77  & 0.06  \\
 X centroid 0\farcs1 Pix      & 242.41  & 0.02  & 242.28  & 0.03  & 754.83  & 0.02  & 754.78  & 0.03  \\
 Y centroid 0\farcs1 Pix      & 700.02  & 0.02  & 700.23  & 0.03  & 325.91  & 0.02  & 326.23  & 0.04  \\
 Half-light radius            & 1\farcs200 & 0\farcs006 & 1\farcs350 & 0\farcs058 & 0\farcs199 & 0\farcs013 & 0\farcs222 & 0\farcs023 \\
 Position Angle light         & \M 37\degpt8   & 0\degpt8   & \M 41\degpt8   & 0\degpt2   &  24\degpt7   & 1\degpt5   &  22\degpt6   & 0\degpt5   \\
 Axis Ratio of the light      &   0.68  & 0.01  &   0.65  & 0.03  &   0.84  & 0.03  &   0.73  & 0.07  \\
 Mag. of lensed source        &  25.76  & 0.2   &  26.1   & 0.2   &  27.4   & 0.2   &  27.4   & 0.2   \\
 Source X offset 0\farcs1 Pix &   0.28  & 0.06  &   0.12  & 0.03  &   0.28  & 0.07  &   0.47  & 0.04  \\
 Source Y offset 0\farcs1 Pix & \M 1.24 & 0.07  & \M 1.36 & 0.04  & \M 0.23 & 0.09  & \M 0.55 & 0.03  \\
 Critical Radius arcsec       &  1\farcs508  & 0\farcs008 &   1\farcs511 & 0\farcs005 &   0\farcs648 & 0\farcs010 &   0\farcs646 & 0\farcs006 \\
 Axis Ratio Lens mass         &  0.395  & 0.018 &   0.370 & 0.006 &   0.64  & 0.04  &   0.369 & 0.011 \\
 Source half-light            &  0\farcs016  & 0\farcs009 &   0\farcs018 & 0\farcs010 &   0\farcs053 & 0\farcs013 &   0\farcs031 & 0\farcs025 \\
 Lens Rotation                &  8\degpt5 & 0\degpt8 &  12\degpt8   & 0\degpt2   &   0\degpt00  & Fixed &   0\degpt00  & Fixed \\
\tableline
\end{tabular}
\vskip 0.5truecm
\end{center}
\end{table*}

\section{THE LENS MODEL}

In this letter we will limit the model analysis to gravitational lenses
described by singular isothermal elliptical potentials (Kormann,
Schneider \& Bartelmann 1994) which provide a sufficiently accurate
representation for our purposes. We adopt their definitions and 
notations in this letter.

Since we had already developed software for 2-dimensional `disk $+$
bulge' decomposition of MDS galaxy images, we used the same procedure
with a slight modification to do `bulge $+$ gravitational lens'
decomposition of the observed light distribution. This procedure
iteratively converges simultaneously on the maximum likelihood model for
the lensing elliptical galaxy and the source, so as to produce the
observed lensing galaxy and the configuration of images from the lensed
source.

For the light distribution of the elliptical, we used the seven
parameters that we routinely use for the analysis of a galaxy image with
a single component: local sky, location (x,y), total magnitude,
half-light radius, axis ratio and position angle. The lensed source was
defined by four parameters: location (sx,sy) relative to the elliptical,
and intrinsic magnitude and half-light radius. Since the morphology of
the source was very poorly constrained, it was assumed to be bulge-like
with spherical symmetry.  The lens is defined by the critical radius
($\xi_0$) and the axis ratio of the mass distribution.  An optional
parameter is used to measure the difference in the orientation between
the mass and the observed light of the lens.

Given a set of model parameters, we generate 2-dimensional images for
the elliptical and the source galaxies.  The expected configuration of
the lensed images is ray-traced by numerical integration.  To improve
the accuracy of the numerical integration, the image is evaluated using
a pixel size smaller than the real one, such that the image fills a 64
pixel square array. The elliptical lens and the lensed source images are
then convolved with the adopted HST WFPC2 point spread function from
TinyTim (Krist 1994).  The convolved image is spatially integrated to
the WFC pixel size of 0\farcs1 and is then compared with the observed
galaxy image. The likelihood function is defined as sum over all pixels
of the logarithm of the probability of the observed value, assuming that 
it has a Gaussian error distribution with respect to the model.  This is
similar to a weighted $\chi^2$, and is then minimized using a
quasi-newtonian method.

\section{MAXIMUM LIKELIHOOD ESTIMATES}

The results of the maximum likelihood estimates and errors are
summarized in Table 1 and the corresponding images are shown in Figure~1
for both of the galaxies analyzed.  The rms errors are estimated from
the covariance matrix which is derived by inverting the Hessian at the
maximum of the likelihood function.  Note that the model parameters
fitted independently to the F814W and F606W band images are very similar
and well within the statistical errors and expected variations due to
color. Since the fainter lensed images are blue and better resolved in
F606W, the error estimates are smaller in this filter, even though the
image exposure times in the F814W filter are 50\% longer.

The critical radii of HST14176+5226 and HST12531+2914 are 1\farcs5 and
0\farcs6 , in each case larger than the half light radii of 1\farcs2 and
0\farcs2 for these galaxies respectively. The lensed image components
are more distant from the centroid of the lensing elliptical galaxy
along the minor axis because the deflection is proportional to the
gradient of the potential.  The ratio in the distance between components
at opposite ends of the cross is equal to the inverse axis ratio of the
potential. The distance of the intrinsic source from the centroid of the
lens needs to be less than about 0.15 of the critical radius for the
creation of a quadruple image. The impact parameters for these two
objects are about 0.08 and 0.09 of the critical radius.

In both cases the source has a half-light radius which is smaller than a
WFC 0\farcs1 pixel. However, the fact that the maximum likelihood
estimate converged to a small but finite intrinsic source half-light
radius appears to indicate that it is extended. We find such a value for
the half-light radius to be similar to those of field galaxies at the
magnitude of the lensed source (see Casertano \etal 1995, Im \etal
1995), i.e. within the range $0\farcs04-0\farcs40$.

\begin{figure}[bt!]
\psfig{file=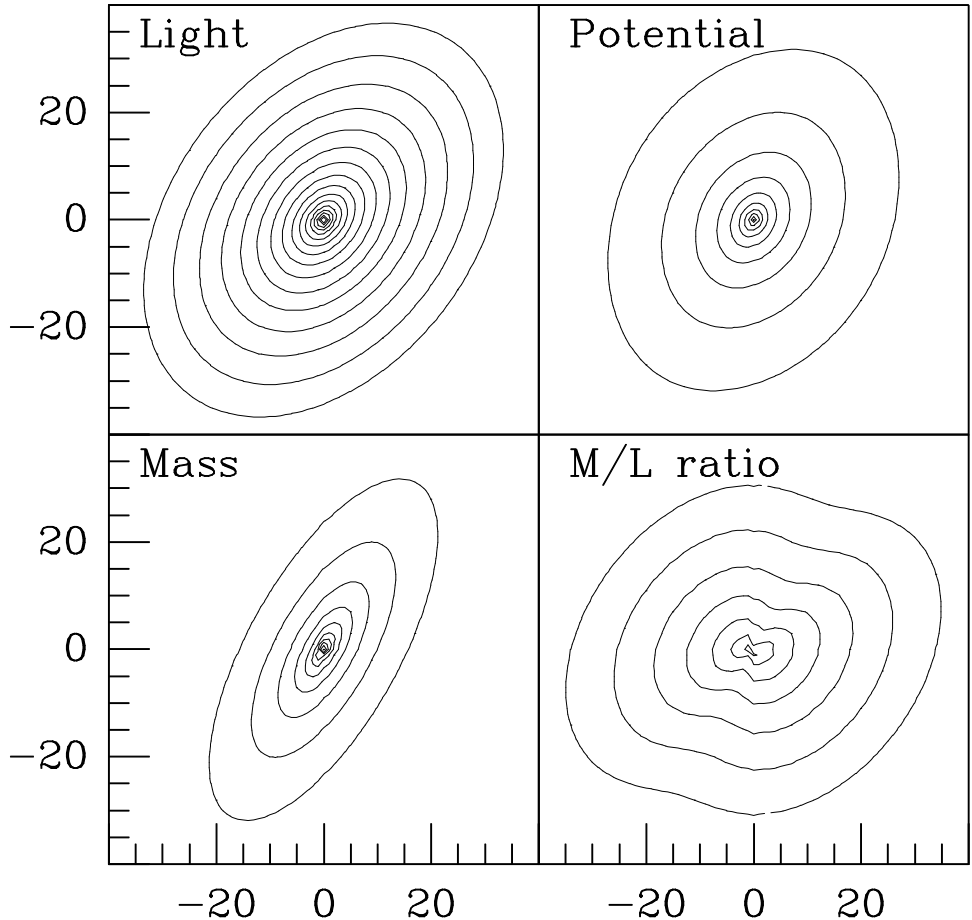,width=5.0in}
\caption{The observed light distribution of HST14176+5226, the
maximum likelihood potential according to the elliptical mass model, and
the resulting Mass/Light distribution, which is rounder (axis ratio=0.82) and
increases radially outwards.  All the contours have been drawn at
constant 0.5 mag intervals, and units are 0\farcs1 WFC pixels. } 
\end{figure}

A significantly better fit (99.9\%) is obtained for HST14176+5226 by
relaxing the constraint that the orientation of the lens mass is the
same as that of the observed light. The difference is practically
independent of the model used, and is 8\degpt5$\pm$0\degpt8 in F814W and
12\degpt9$\pm$0\degpt2 in F606W.  Such a small rotation might be
expected from projection effects (see Franx, Illingworth \& de Zeeuw 1991). 
As shown in Figure 2, the axis ratio
(0.68) of the light from the elliptical galaxy HST14176+5226 is
practically the same as that of the potential (0.74).  The inferred
elliptical mass distribution is significantly flatter than this (0.40).

Using the observed V\M I colors, apparent magnitudes and half-light
radii, we can estimate the redshift for each of the lens elliptical
galaxies to be $z(D_d)=0.7\pm 0.1$ (see, e.g., Connolly \etal 1995).
Using these redshifts, the luminosities of the elliptical galaxies can
be estimated for the plausible range of evolutionary and K-corrections
and the value of $\Omega$, and thus the velocity dispersion $\upsilon$
from the relation of Faber \& Jackson (1976).  We find that the observed
critical radii $\xi_0=4\pi (\upsilon /c)^{2} D_d D_{ds}/D_s$, are more
probable for $\Omega < 1 $ for a source redshift smaller than $z(D_s) <
3 $ (Guhatakurta, Tyson \& Majewski 1990).  For more detailed
discussions, see Im \etal (1995).

\section{CONCLUSIONS}

We have discovered two examples of `Einstein-cross' gravitational lenses
in HST survey data, one in the MDS data and one in the archived
Groth-Westphal GTO survey.  The lens configurations show that the stars
responsible for the light distribution of these elliptical galaxies are
a trace population following the gravitational potential. The Mass to Light 
ratio increases radially outwards.

These represent the first discoveries of lenses using the high
resolution of HST - indeed, apart from the exceptional original Einstein
cross discovered by Huchra \etal (1985), they represent the first
discoveries of field-galaxy gravitational lenses via the systematic
study of optical images.  They are a new class of gravitational lens
candidates in which the cosmologically distant lens is a relatively
bright elliptical galaxy with well understood properties.  If a
significant sample could be found and observed spectroscopically for
redshifts, they will be very useful cosmological probes.

These objects would have been very difficult discoveries from the ground
except under conditions of excellent seeing. Indeed, CCD frames taken at
the KPNO 4m prime focus (Connolly 1995) do not show the cross objects
flanking the elliptical galaxy HST14176+5226.  We have not as yet
observed these galaxies spectroscopically.  The redshift of the lensed
components in HST12531\M2914 is probably a challenging observation for
the Keck telescope in excellent seeing.

From the observed numbers of bright elliptical galaxies observed in the
GTO survey strip (300 to I=22), the numbers of faint objects in the
fields (8000 to I=26), and the expected cross-sections, we estimate
that we should find one quadruple lens in every $20-30$ WFPC2 fields
surveyed. The number that has been discovered so far is therefore
consistent with our expectations. An on-going systematic and careful
inspection of the MDS fields, looking very specifically for possible
gravitational lens candidates in the shallower MDS and GTO parallel
fields is in progress in order to expand the sample. As further MDS data
are taken in Cycle 5 and subsequent cycles, they will be examined for
similar spectacular lenses, and also for more common lenses consisting
of arcs or two or three components, to obtain a statistically
representative sample of HST gravitational lens candidates.

This paper is based on observations with the NASA/ESA Hubble Space
Telescope, obtained at the Space Telescope Science Institute, which is
operated by the Association of Universities for Research in Astronomy,
Inc., under NASA contract NAS5-26555.  The Medium-Deep Survey is funded
by STScI grant GO2684. We gratefully acknowledge Lyman Neuschaefer for
help with the MDS pipeline and many helpful discussions, the anonymous
referee from many useful suggestions, and {\it et tu} Broadhurst.

\end{document}